\documentclass[multphys,vecphys]{svmult}

\usepackage{makeidx}         % allows index generation
\usepackage{graphicx}        % standard LaTeX graphics tool
\usepackage{multicol}        % used for the two-column index
\usepackage[bottom]{footmisc}% places footnotes at page bottom

\graphicspath{{./Input/}}

\makeindex             % used for the subject index
\newcommand{\Vline}{\ensuremath{V_{\rm line}}}
\newcommand{\Vcont}{\ensuremath{V_{\rm cont}}}

\newcommand{\microns}{\ensuremath{\mu\mbox{m}}}

\newcommand{\degrees}{\ensuremath{\mbox{deg}}}
\newcommand{\yr}{\ensuremath{\mbox{yr}}}

\newcommand{\kms}{\ensuremath{\mbox{km\,s}^{-1}}}

\newcommand{\Msun}{\ensuremath{M_\odot}}
\newcommand{\MsunPyr}{\ensuremath{\Msun\,\yr^{-1}}}
\newcommand{\Ha}{\ensuremath{\mbox{H}\alpha}}
\newcommand{\Hb}{\ensuremath{\mbox{H}\beta}}
\newcommand{\Brg}{\ensuremath{\mbox{Br}\gamma}}
\newcommand{\lBrg}{\ensuremath{2.1656\,\microns}}
\newcommand{\mwc}{\ensuremath{\mbox{MWC}\,297}}

\begin{document}

\title*{Disentangling the wind and the disk in the close surrounding of
  the young stellar object MWC297 with AMBER/VLTI}
\titlerunning{Wind and disk resolved in \mwc\ with AMBER/VLTI}
\author{F. Malbet\inst{1}
  \and M. Benisty\inst{1}
  \and W.J. de Wit\inst{1}
  \and S. Kraus\inst{2}
  \and A. Meilland\inst{3}
  \and F. Millour\inst{1,4}
  \and E. Tatulli\inst{1}
  \and J.-P. Berger\inst{1}
  \and O. Chesneau\inst{3}
  \and K.-H. Hofmann\inst{2}
  \and A. Isella\inst{5}
  \and R. Petrov\inst{4}
  \and T. Preibisch\inst{2}
  \and P. Stee\inst{3}
  \and L. Testi\inst{5}
  \and G. Weigelt\inst{2}
  \and the AMBER consortium
}             
\authorrunning{F. Malbet et al.}
\institute{
% 1
  Laboratoire d'Astrophysique de Grenoble, UMR 5571 Universit\'e Joseph
  Fourier/CNRS, BP 53, F-38041 Grenoble Cedex 9, France
% 2
  \and Max-Planck-Institut f\"ur Radioastronomie, Auf dem H\"ugel 69,
  D-53121 Bonn, Germany
% 3
  \and Laboratoire Gemini, UMR 6203 Observatoire de la C\^ote
  d'Azur/CNRS, BP 4229, F-06304 Nice Cedex 4, France
% 4
  \and Laboratoire Universitaire d'Astrophysique de Nice, UMR 6525
  Universit\'e de Nice/CNRS, Parc Valrose, F-06108 Nice cedex 2, France
% 5
  \and Osservatorio Astrofisico di Arcetri, Istituto Nazionale di
  Astrofisica, Largo E. Fermi 5, I-50125 Firenze, Italy
}
\maketitle

\abstract{The young stellar object \mwc\ is a B1.5Ve star exhibiting
  strong hydrogen emission lines. This object has been observed by the
  AMBER/VLTI instrument in 2-telescope mode in a sub-region of the K
  spectral band centered around the \Brg\ line at 2.1656\microns. The
  object has not only been resolved in the continuum with a visibility
  of $0.50\pm0.10$, but also in the \Brg\ line, where the flux is
  about twice larger, with a visibility about twice smaller
  ($0.33\pm0.06$).  The continuum emission is consistent with the
  expectation of an optically thick thermal emission from dust in a
  circumstellar disk. The hydrogen emission can be understood by the
  emission of a halo above the disk surface. It can be modelled as a
  latitudinal-dependant wind model and it explains the width, the
  strength and the visibibility through the emission lines.  The AMBER
  data associated with a high resolution ISAAC spectrum constrains the
  apparent size of the wind but also its kinematics.}
\section{Introduction}
\label{sect:intro}

Pre-main sequence stars in the intermediate mass range, called Herbig
Ae and Be stars (HAeBe), are observed to be surrounded by
circumstellar material. It reveals itself by discrete emission lines
and by continuous excess emission in the spectral energy distribution
(SED). The spatial distribution of this material however has been
subject to debate, where both geometrically flat disk models and
spherically symmetric envelope models can reproduce the observed SED.

The geometry of circumstellar material near HAeBe stars seems to
differ between the early-type and late-type members of the group,
which is not surprising given the increasing interaction between star
and disk at the earlier type stars.  For the HAe stars a successful
working model exist, while on the other hand, a disk structure near
the HBe stars and their intricate star-disk interactions still escape
a good understanding. In this study we present high spatial
resolution, intermediate spectral resolution interferometric
observations with AMBER of the early-type Herbig Be star \mwc.  This
star displays a strong emission line spectrum corresponding to a
B\,1.5Ve spectral type. The rather well determined stellar parameters
\cite{Drew1997} and its high NIR luminosity render this star the
perfect target to investigate in detail the geometry of the
circumstellar material near the early type HAeBe stars.

\section{Observations}

\mwc\ has been observed during the second night of the first
commissioning run of the AMBER instrument on the UT2-UT3 baseline of
the \emph{Very Large Telescope Interferometer} (VLTI).  AMBER is the
VLTI beam combiner in the near-infrared \cite{Pet2003}.  The
instrument is based on spatial filtering with fibers and spatial beam
combination along one dimension. The interferometric beam is
anamorphized perpendicular to the fringe coding in order to be
injected into the slit of a spectrograph. \mwc\ has been measured in
the [2100,2230nm] spectral range with 1500 spectral resolution.  

The results for the line visibilities are relatively consistent with
all data reduction methods \cite{Mil2004, Ohn2003}, while this is not
the case for the continuum visibilities. Besides the continuum
visibilities in the K band has already been measured by other
instruments like IOTA and PTI and therefore the important result is
the line visibility.

\begin{figure}[t]
  \centering
  \parbox[c]{0.415\hsize}{\vspace*{1em}\includegraphics[height=\hsize]{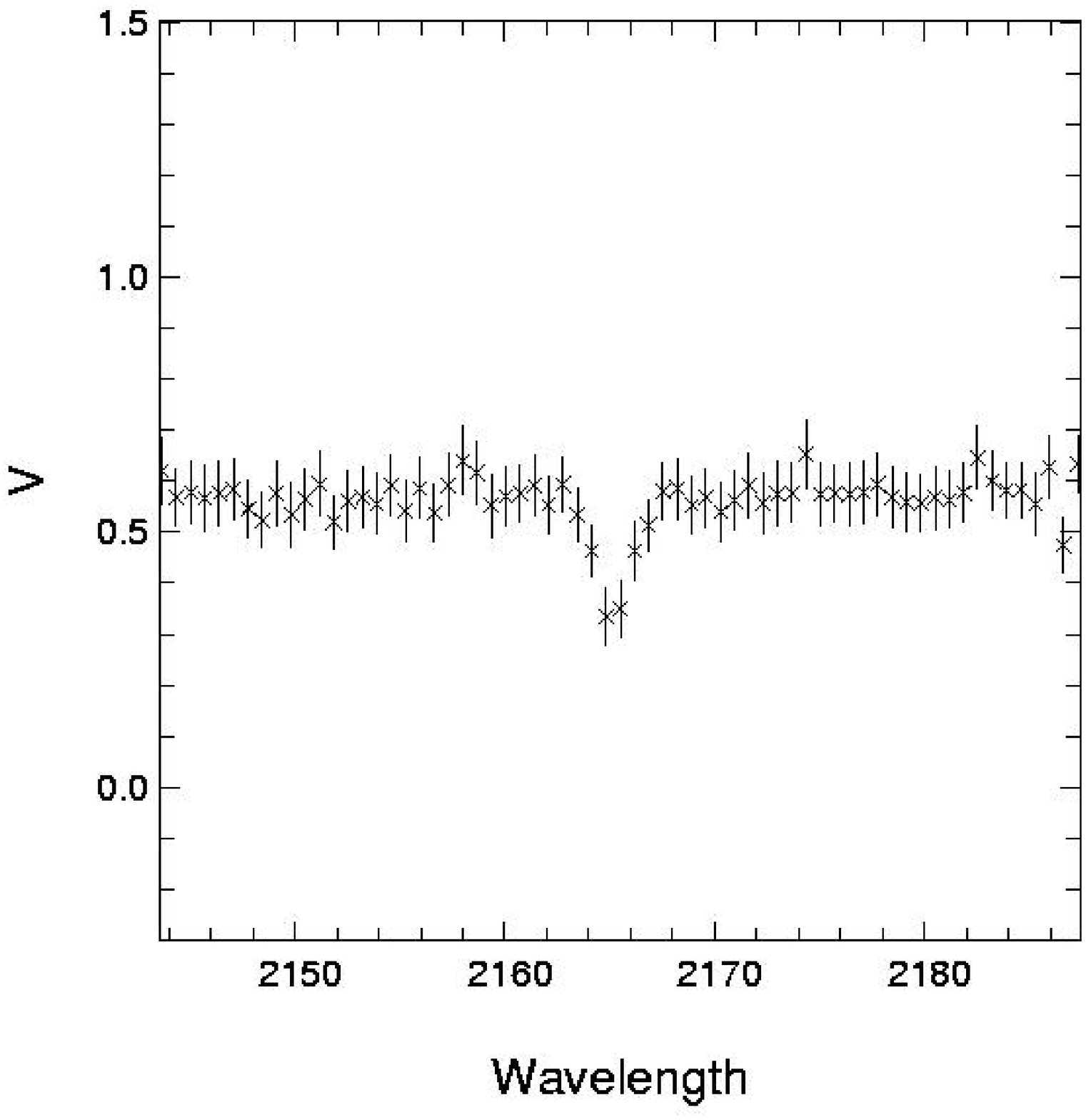}}
  \parbox[c]{0.4\hsize}{\includegraphics[height=\hsize]{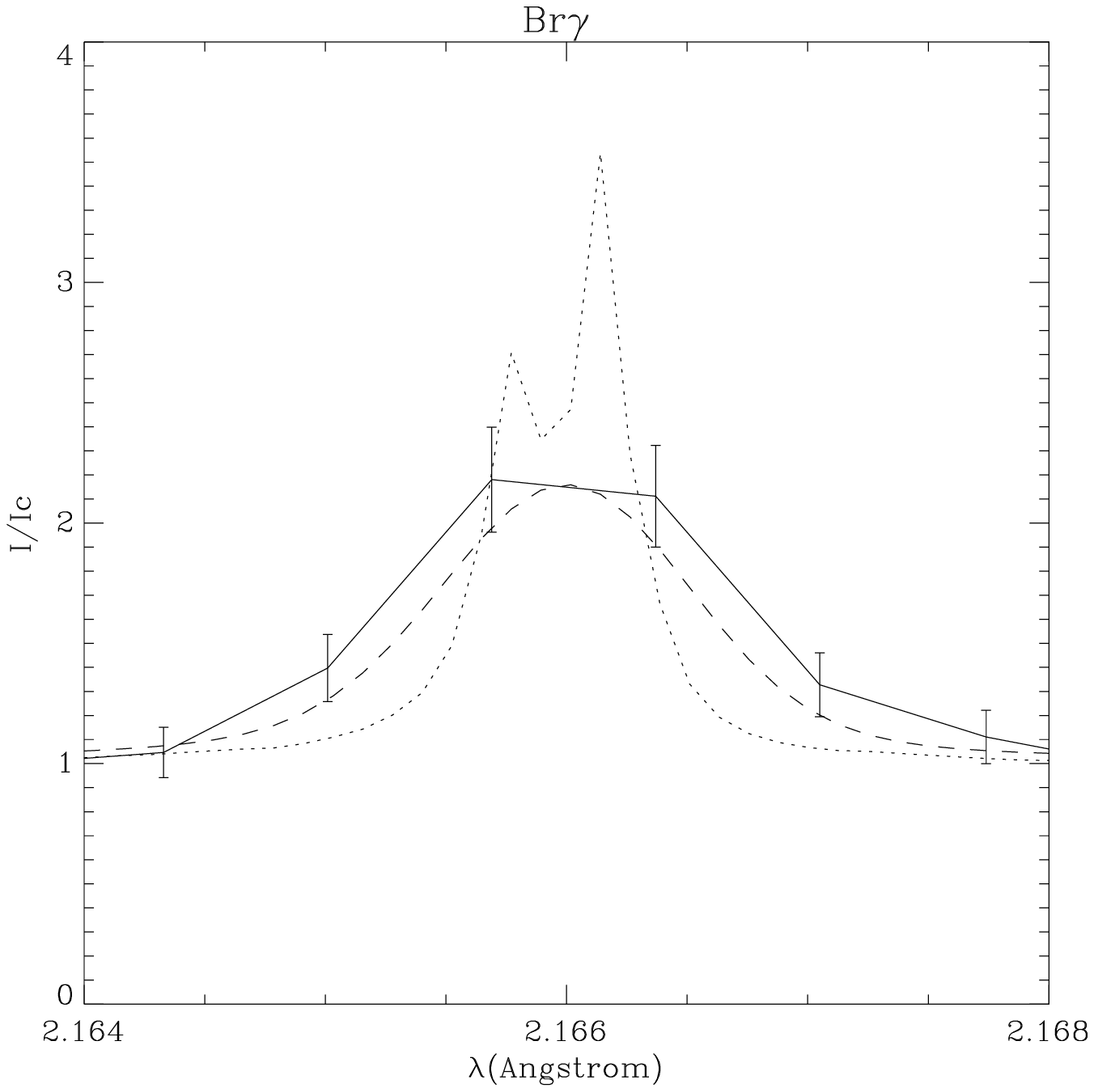}}
  \caption{Left: spectral dependence of the \mwc\ visibilities. Right:
    comparison of \Brg\ observed with AMBER (full line) and ISAAC
    (dotted line).  The dashed line corresponds to the ISAAC spectrum
    convolved at the AMBER spectral resolution.}
\label{fig:results}
\end{figure}

Left part of Fig.~\ref{fig:results} shows the variation of the visibility with
wavelength. 
The continuum visibilities correspond to an average of
$\Vcont=0.50\pm0.10$ and the line visibility to a value of
$\Vline=0.33\pm0.06$.

\mwc\ was also observed with ISAAC at the ESO VLT UT1 telescope (see
right part of Fig.~\ref{fig:results}) in the short wavelength medium
resolution mode ($\mathcal{R}=8900$) at the \Brg\ wavelength.
Broad-band photometric data were collected from the litterature
\cite{Drew1997,Pez1997,Man1994}.  Existing interferometric data for
\mwc\ consist of IOTA H-band \cite{MST2001} and PTI K-band
\cite{Eis2004} continuum data.

\section{Modeling}
\label{sect:models}

We tried to model the large body of interferometric, spectroscopic and
photometric data that exists for \mwc. The modeling is done by
applying two different codes, an optically thick disk one and a
stellar wind one. The disk code is designed to model the continuum
radiation, whereas the stellar wind code reproduces the strong
emission lines.
\begin{figure}[t]
  \centering
  \includegraphics[width=0.7\hsize]{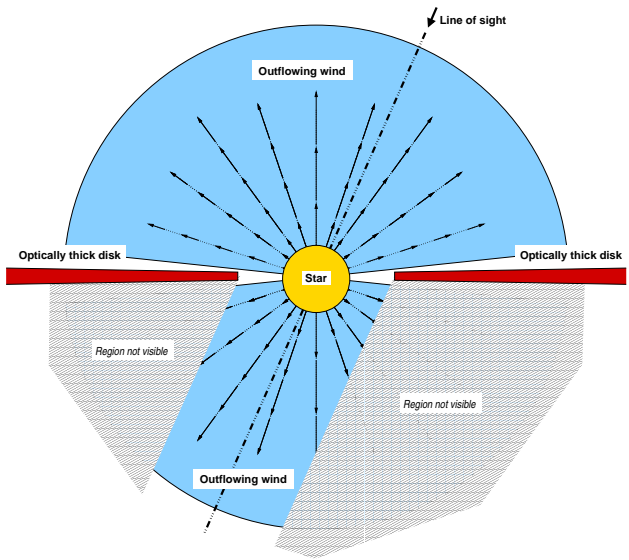}
  \caption{Sketch of the model including an optically thick disk and
    an outflowing wind (edge-on view). The receding part of the wind
    is only partly visible because of the screen made by the optically
    thick disk.}
  \label{fig:model}
\end{figure}
Figure \ref{fig:model} represents a sketch of the combined model, where the
optically thick disk and the outflowing wind are spatially independent.

\subsection{Continuum radiation: optically thick disk}
\label{sect:disk}

\begin{figure*}[t]
{
  \begin{center}
    \includegraphics[width=0.4\hsize]{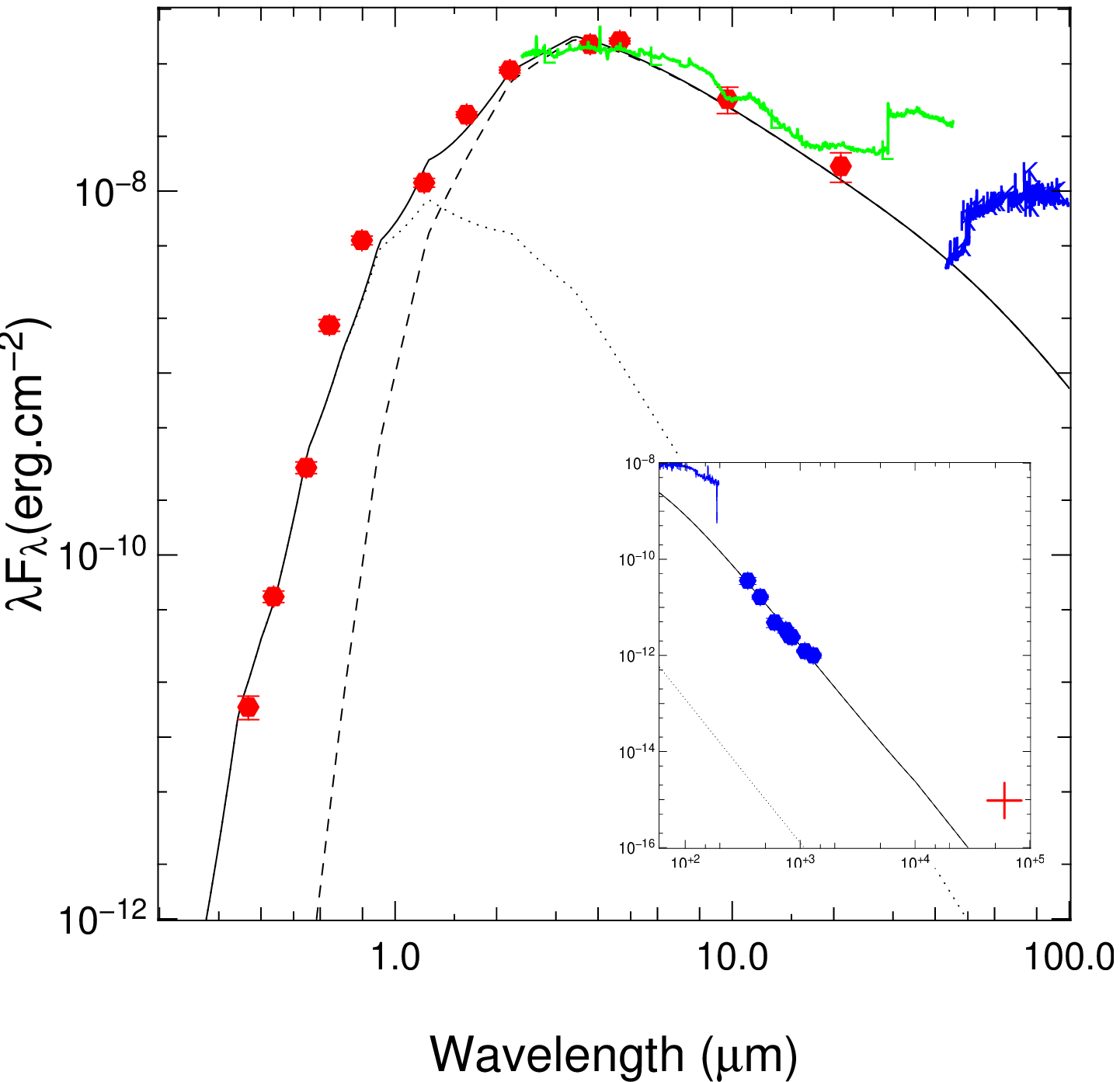}
    \includegraphics[width=0.38\hsize]{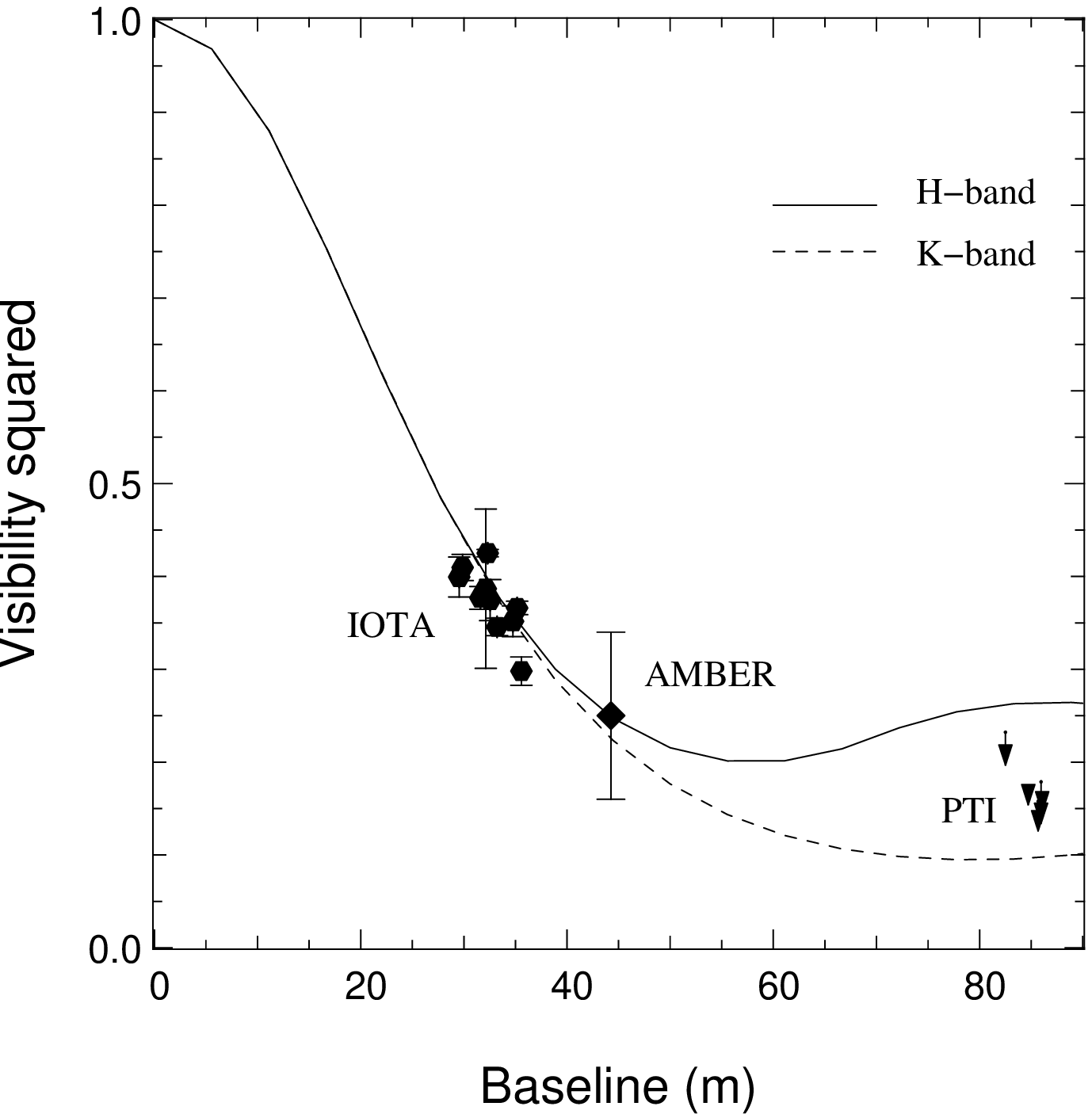}
    \caption{Result from the optically disk model. {\it Left panel:}
      observed and modeled SED for \mwc. The full dots are the
      continuum measurements \cite{Pez1997}, also included
      are the ISO SWS/LWS spectra.  Dotted line is the star, dashed
      line the accretion disk, and the full line the resulting total
      flux of the model. {\it Right panel:} resulting best-fit model
      radial visibilities compared with AMBER, IOTA and PTI observed
      continuum visibilities. Full line and IOTA data are in the
      $H$-band, dashed line and AMBER/PTI are in the $K$-band. PTI
      values are upper limits.}
\label{fig:sed}
\end{center}
}
\end{figure*}

The disk model \cite{MB1995,Mal2005} consists in an axisymmetric
radial analytic disk structure which is heated both by stationary
accretion and stellar irradiation. The disk is in hydrostatic
equilibrium, non self-gravitating and the accretion flux is following
the standard power law for a viscous disk. The emitted continuum flux
is produced by the emission of optically thick but geometrically thin
black-body radiating rings. It produces an SED, and, its spatial
distribution can be Fourier transformed to obtain interferometric
visibilities.

We probed the sensitivities of these fits by varying the central star
parameters, according to the uncertainties given by Drew et al.
\cite{Drew1997}.  They derived half a spectral subtype uncertainty,
and a distance error of 50\,pc. The mass accretion rate is far
from well determined. If the central star would be of type B2 at a
distance of 200\,pc, the required mass accretion rate is between 0 and
$10^{-6}\,\MsunPyr$.

\subsection{Emission lines: optically thin outflowing wind}
\label{sect:wind}

In our model, the emission lines are produced in a circumstellar gas
envelope. In order to model this line profile and the corresponding
visibilities, we have used the SIMECA code \cite{SA1994,Stee1995}.
The solutions for all stellar latitudes are obtained by introducing a
parametrized model constrained by the spectrally resolved
interferometric data.

Since the SIMECA code has originally been developed to model the
circumstellar environment of classical Be stars, we had to modify the
code in order to interface SIMECA with the optically thick disk model
described previously. We have implemented three changes:
\begin{enumerate}
\item The wind is no longer computed from the equator to the pole, but
  the computation occurs in a bipolar cone defined by a minimal angle
  allowing the disk to be present (see sketch in
  Fig.~\ref{fig:model}). We used an minimum angle of 4 degrees.
  The equatorial terminal velocity corresponds therefore to the
  terminal velocity at this minimal angle from the equatorial plane at
  the interface between the accretion disk and the stellar wind.
\item The disk hides the receding part of the wind. In
  Fig.~\ref{fig:model}, the part of the wind which is not visible from
  the observer is not taken into account in the outgoing flux.
\item Although the disk emission contributes less than $1\%$ compared
  to the star flux in the visible (i.e.\ also in the \Ha\ and \Hb\
  lines) and can be neglected, at \lBrg\ the disk emission is 6.4 times
  larger than the stellar flux.  This contribution decreases the
  normalized \Brg\ line intensity and also must be accounted for
  in the computation of the visibilities.
\end{enumerate}

\begin{figure*}[t]
  \centering
  \includegraphics[width=0.4\hsize]{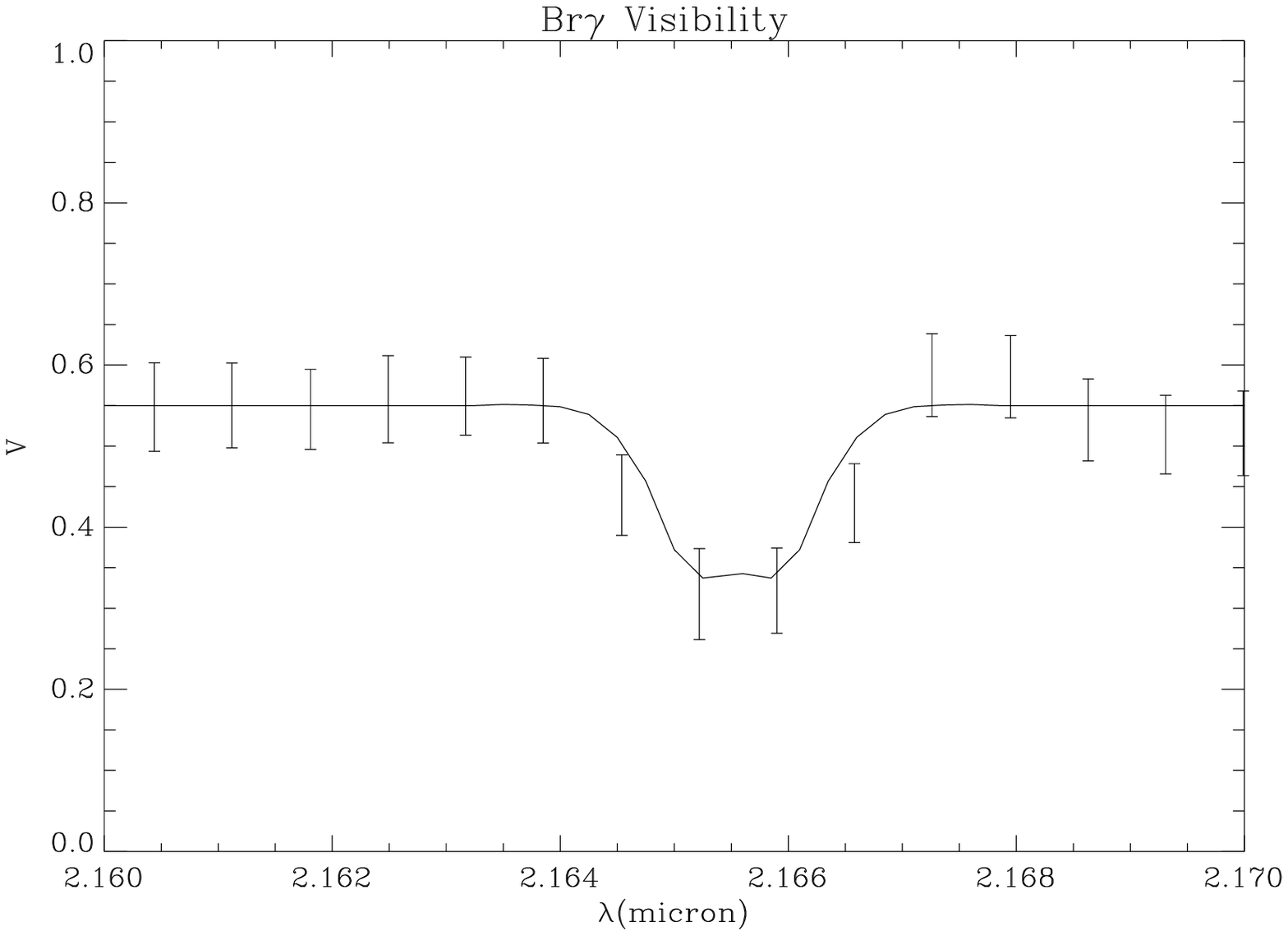}
  \hfill
  \includegraphics[width=0.4\hsize]{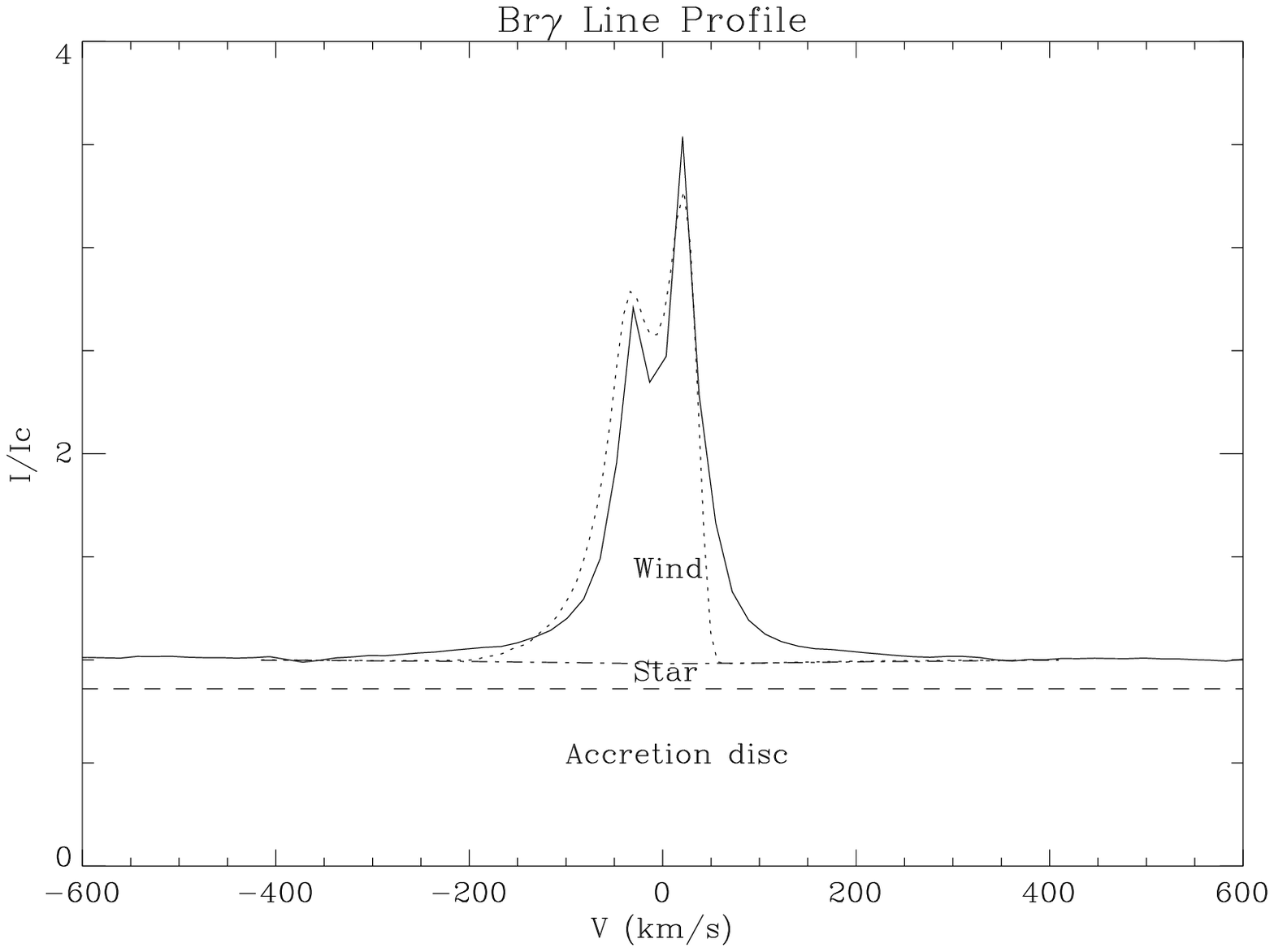}
  \caption{{\it Left panel:} The visibility observed with AMBER
    (points with error bars) and the modeled outflowing wind model
    (full line). {\it Right panel:} double peaked \Brg\ profile
    observed by ISAAC (full line) and modeled with the outflowing wind
    model (dotted line). We have also plotted the cumulative
    contribution of the accretion disk (dashed line) and of the star
    (dash-dot).}
    \label{fig:spectmodel}
\end{figure*}
We find a successful simultaneous fit to the \Ha, \Hb\ and \Brg\ line
profiles compatible with the observations (see
Fig.~\ref{fig:spectmodel}) and the outflowing wind model reproduces
the AMBER measured drop in visibility across the \Brg\ line.

We are able to reproduce quite well the shape of the \Ha\ and \Hb\ 
lines and the double peaked emission of the \Brg\ line. The peak
asymmetry of the \Brg\ line is also reproduced thanks to
the introduction in the SIMECA code of the opacity of the disk (point
2 of SIMECA modifications). Nevertheless the agreement is not perfect
in the red wing of the profile probably due to our ad-hoc way of
interfacing of the wind and the disk.

\section{Discussion}
\label{sect:discussion}

The modeling presented in the previous section, although rather
successful, brings new questions on the physics of the circumstellar
environment of intermediate-mass young stars. 

\begin{figure*}[t]
  \centering
  \includegraphics[width=0.3\hsize]{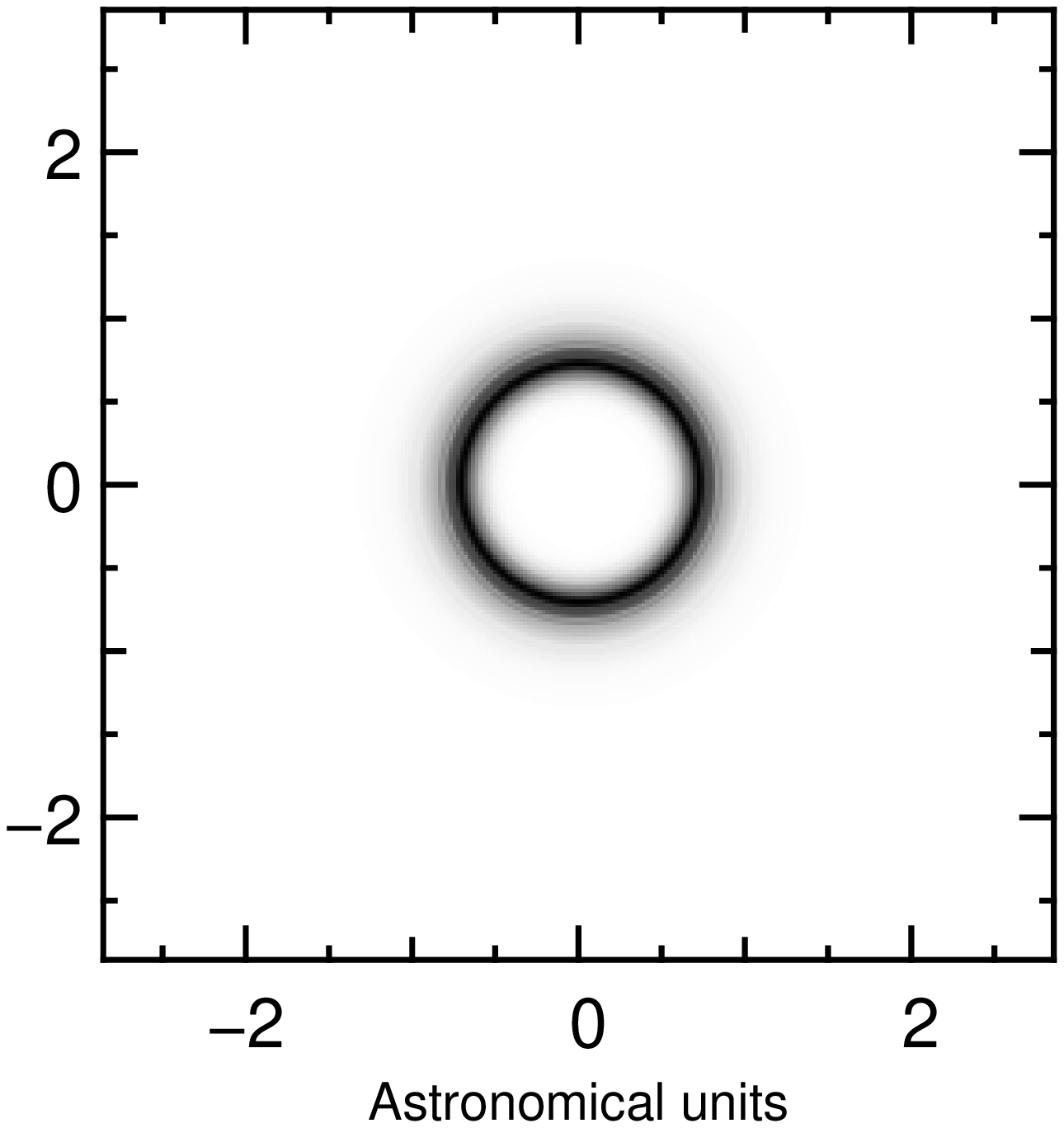} 
  \hfill
  \includegraphics[width=0.3\hsize]{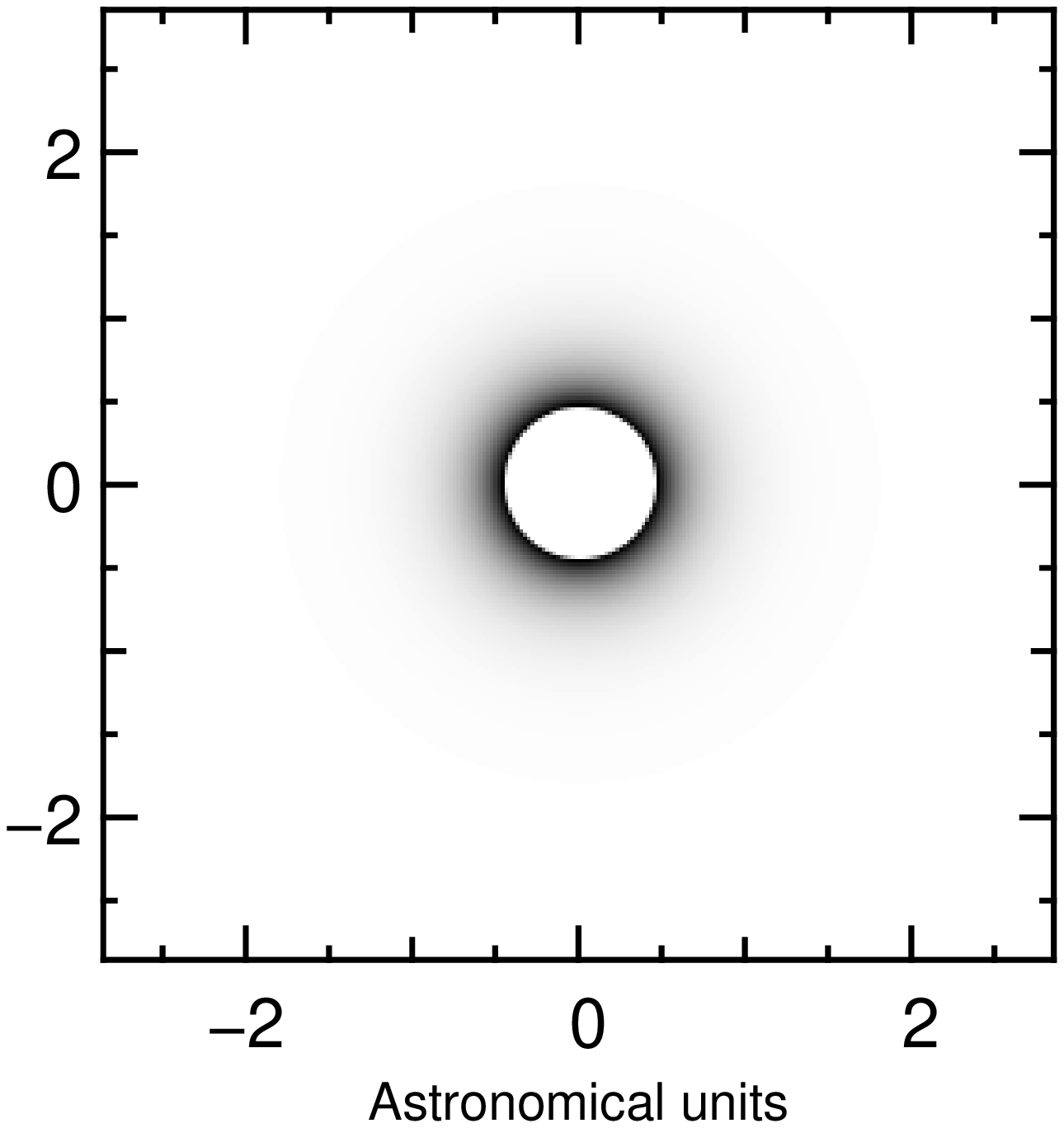}
  \hfill
  \includegraphics[width=0.3\hsize,height=0.32\hsize]{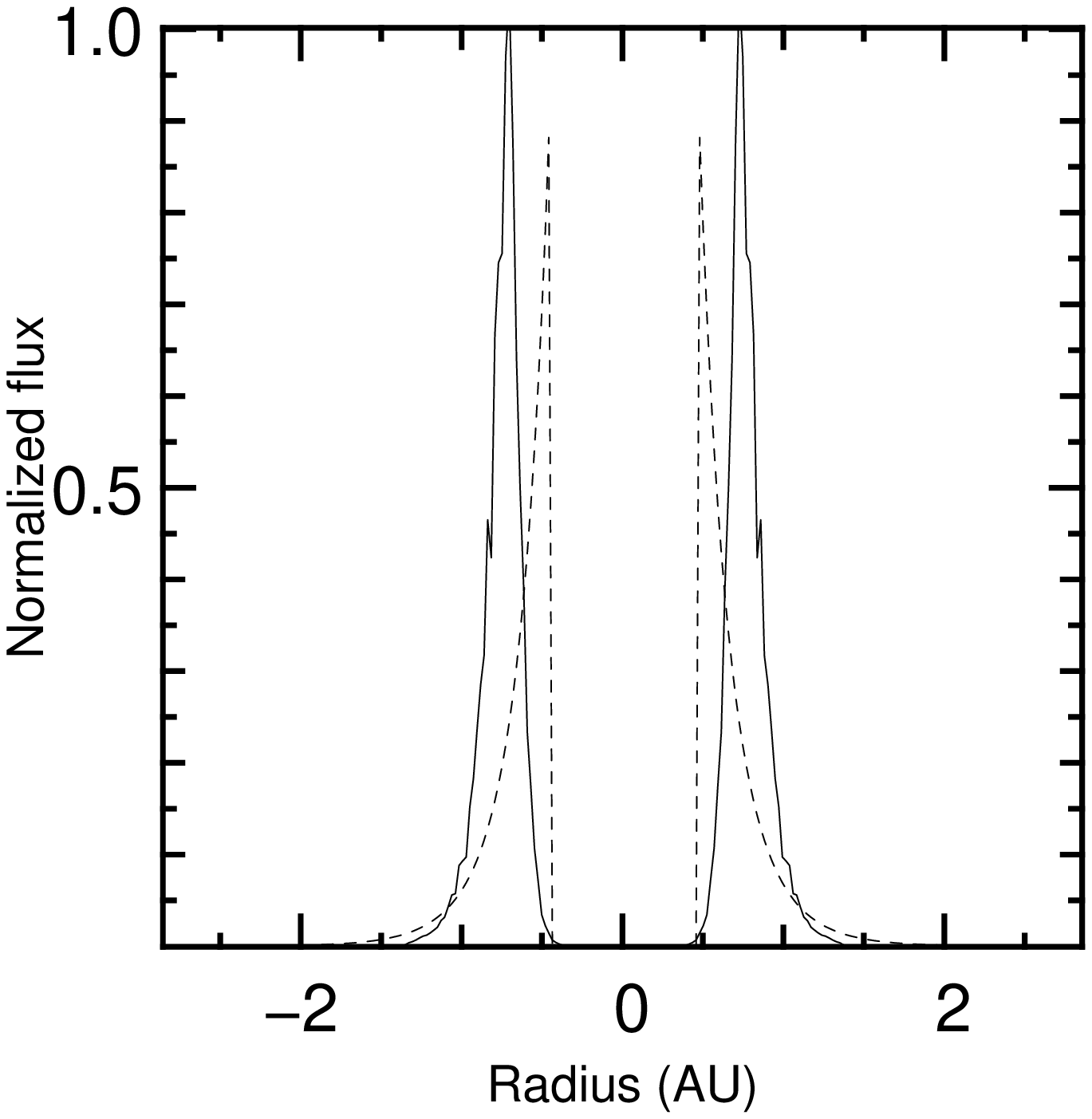} 
  \caption{Pole-on intensity maps of the wind \Brg\ emission (left
    panel) and of the $K$-band disk continuum emission (center panel).
    Right panel shows a radial cut of these intensity maps with the 
    \Brg\ wind in solid line and the continuum disk in dashed line.}
  \label{fig:intmaps}
\end{figure*}
Figure \,\ref{fig:intmaps} shows the pole-on intensity maps of the disk
model in the continuum and of the wind in the \Brg\ line, as well as
their respective intensity profile. This is a graphical explanation of
the visibilities observed by AMBER: the wind angular extension in the
\Brg\ line is larger than the disk apparent size and therefore the
visibility is smaller within the line.

Can the result obtained with these observations constrain the nature
of the wind? We recall that in protoplanetary disks, two main classes
of disk wind models have been proposed depending of the geometry of
the magnetic field lines: the disk wind \cite{BP1982,CF2000} and the
X-wind \cite{Shu1994}.  We cannot conclude with the present state of
data since we are unable to recover a precise kinematical map of the
wind. More resolution with AMBER will help to answer the question,
especially using the 10000 spectral resolution mode.

Also the modeling presented in this paper does a reasonably good job in
reproducing nearly all the observational data and produces fiducial
physical parameters for the circumstellar environment of \mwc. However,
we derive an inclination of $\sim 20\,\degrees$ for the system, which
is not consistent with a near edge-on orientation as proposed by
\cite{Drew1997}. The later is inferred from the photospheric lines
that indicate a $350\,\kms$ projected rotational velocity. An
inclination of $20\,\degrees$ would lead to a rotation above the
break-up velocity.

In conclusion, we can claim that the models of disk and wind are
compatible and are probably very close from the reality. A complete
and self-consistent modeling of the environment is out of the scope
of the paper but would allow to better constrain the relationship
between the disk and the wind at least from the observational point of
view.

\section{Conclusion} 
\label{sect:summ}

We have presented first spatially resolved observations of the disk /
wind interaction in the young stellar system \mwc\ with the VLT
interferometer equipped with the instrument AMBER. We have observed
that the continuum visibility in the $K$-band drops from 0.50 to 0.33
in the \Brg\ emission line of \mwc. The spectrum obtained with AMBER
is consistent with a double peaked spectrum observed with ISAAC on the
VLT, where the peaks are roughly separated by about $60\,\kms$.

We have successfully modeled the circumstellar environment of \mwc\
using an optically thick geometrically thin disk and an outflowing
stellar radial wind having a increasing outflowing velocity starting
from the surface of the disk up to the pole. This combined model is
able to reproduce many observational features like the shape of the
SED over more than three decades of the wavelengths, the continuum
visibilities obtained not only by AMBER but also by other infrared
interferometers like IOTA and PTI, the spectral visibilities in the
\Brg\ emission line as well as the \Ha\, \Hb\ and \Brg\ line profiles.

We are not yet able to constrain the exact nature of the wind and the
type of connection with the disk, but we expect that future data obtained
with AMBER at a higher spectral resolution will give new kinematical
information on this interesting and intriguing region.

\bibliographystyle{springer}
\bibliography{malbet-mwc297}

\end{document}